\documentstyle[aps,prb,epsf,twocolumn,floats]{revtex}

\newcommand{\Irect}{I_{\rm rect}}

\newcommand{\Vrect}{V_{\rm rect}}
\newcommand{\Vpump}{V_{\rm pump}}

\begin{document}
\draft
\title{Rectification of displacement currents
in an adiabatic electron pump}

\author{P. W. Brouwer}
\address{Laboratory of Atomic and Solid State Physics,
Cornell University, Ithaca, NY 14853-2501\\
{\rm (\today)}
\medskip ~~\\ \parbox{14cm}{\rm
Rectification of ac\ displacement currents generated by
periodic variation of two independent gate voltages of a 
quantum dot can lead to a dc\ voltage linear in the 
frequency $\omega$. The presence of this rectified displacement
current could account for the magnetic field symmetry observed 
in a recent measurement on an adiabatic quantum electron 
pump by Switkes {\em et al.} [Science {\bf 283}, 1905 
(1999)].\medskip }}
\maketitle 

\narrowtext

In a recent publication, Switkes {\em et al.}\ reported
measurements on an electron pump that uses periodic
changes of the electronic wavefunction 
as the pumping mechanism.\cite{switkes}
This ``adiabatic quantum electron pump'' consists of a
quantum dot coupled to two electron reservoirs via
ballistic point contacts. The shape of the quantum dot
can be controlled by two independent gate voltages $X_1$
and $X_2$. The shape changes are small enough that they
only affect the quantum mechanical wavefunction
of electrons in the quantum dot, not their classical 
trajectories.
Out-of-phase variation of the gate voltages generates a dc\
voltage $\bar V$ across the dot or a dc\ current $\bar I$ 
through the dot, depending on the measurement setup. 

While some features of the observed dc\ voltage $\bar V$
agree well with the theoretical models of an adiabatic quantum 
electron pump,\cite{brouwer,zhou,shutenko} three remarkable 
observations of the experiment could not be explained.
(Following Ref.\ \onlinecite{switkes}, 
I focus on the voltage measurement; the situation of a 
current measurement, which was considered in the
theories of Refs.\ \onlinecite{brouwer,zhou,shutenko}
is discussed later.)
First, for small driving amplitudes $\delta X_{1,2}$, where $\bar V$
is proportional to $\delta X_1 \delta X_2$, the measured dc\
voltage 
is a factor $\sim 20$
higher than the maximum voltage allowed by the
theory for an adiabatic quantum pump in 
the small-amplitude regime.\cite{switkes,brouwer} 
Second, the measured
voltage $\bar V$ is symmetric under reversal of the magnetic
field $B$, while the theory predicts no symmetry relation between
$\bar V(B)$ and $\bar V(-B)$.\cite{shutenko,aleiner,footsymmetry}
(Though implicit in the theory of Ref.\ \onlinecite{brouwer}, 
this was not noted until after publication of the experimental 
findings.)
Furthermore, in the experiment, the magnitude of
the mesoscopic fluctuations of the dc\ voltage decreases
by a factor two upon application of a
time-reversal symmetry breaking magnetic field, 
while the theory has no
difference between $B=0$ and $B\neq0$.\cite{brouwer,shutenko}

In this note, I show that the experimental observations
of Ref.\ \onlinecite{switkes} can be explained if
the device of Ref.\ \onlinecite{switkes} also
serves as a rectifier for an ac\ bias current generated by
a (parasitic) capacitive coupling of the gate voltage to the 
electron reservoirs.\cite{footswitkes} This scenario does not
rule out the possibility that the device of Ref.\ 
\onlinecite{switkes} also pumps electrons. However, to explain 
the experimental data it is sufficient that the rectification 
voltage obscures the true pumped voltage.

\begin{figure}
\epsfxsize=0.95\hsize
\hspace{0.02\hsize}
\epsffile{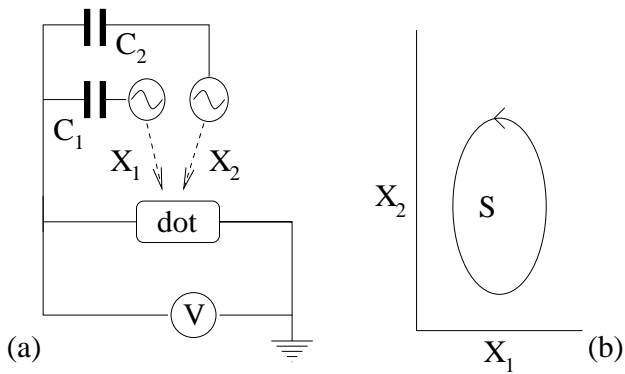}
\medskip

\caption{\label{fig:1} (a) Equivalent electrical circuit for the
experiment of Switkes {\em et al.}\protect\cite{switkes} 
The two ac\ gate voltages $X_1$ and $X_2$ not only control the shape
of the quantum dot (indicated by the dashed arrows), but they are
also coupled to the electron reservoir via stray capacitances 
$C_1$ and $C_2$. The voltage across the quantum dot is measured by 
a current volt meter. (For a current measurement, the volt meter is
replaced by a current meter with resistance $R$.)
(b) Schematic drawing of the contour in 
($X_1,X_2$) space that is traced out by one cycle of the parameters
$X_1$ and $X_2$.}
\end{figure}

An equivalent electrical circuit depicting the experimental
setup for a voltage measurement and including the stray 
capacitances
$C_1$ and $C_2$ between the gates and the electron reservoirs, 
is shown in Fig.\ \ref{fig:1}a. 
At low frequencies, the ac\ gates voltages $X_1$ and $X_2$
generate the current 
\begin{equation}
  I(t) = C_1 {dX_1 \over dt} + C_2 {dX_2 \over dt}
\end{equation}
through the quantum dot.
For slow variations of the gate voltages $X_1$ and $X_2$, the 
voltage $V(t)$ across the quantum dot is
$V(t) = I(t)/G(t)$,
where $G(t)$ is the dc\ conductance of the quantum
dot and the time-dependence of $G$ arises from the time-dependence
of the gate-voltages $X_1$ and $X_2$ controlling the shape of
the quantum dot. The dc\ component $\Vrect$ of the
voltage is found by 
averaging over one period $\tau=2\pi/\omega$,
\begin{eqnarray}
  \Vrect &=& {1 \over \tau} \int_0^{\tau} dt V(t),
\end{eqnarray}
which 
can be rewritten as an integral over the area $S$ enclosed by
the contour in ($X_1,X_2$) space traced out by the gate voltages $X_1$
and $X_2$ in one cycle (see Fig.\ \ref{fig:1}b), 
\begin{eqnarray}
  \Vrect &=& {\omega \over 2 \pi} 
             \int_S dX_1 dX_2 {1 \over G^2} 
             \left(C_1 {\partial G \over \partial
             X_2} - C_2 {\partial G \over X_1} \right).
  \label{eq:I}
\end{eqnarray}
Note that variation of only one gate voltage does not produce 
a dc\ current to first order in the frequency $\omega$.

If the experimental setup is such, that current
is measured, rather than voltage, the 
ac\ gate voltages $X_1$ and $X_2$
serve as an ac\ voltage source
across the quantum dot, $V(t) = R C_1 dX_1/dt + R C_2 dX_2/dt$,
where $R$ is the resistance of the circuit path containing the
current meter (assumed to be
much smaller than the resistance of the quantum dot). 
Averaging over the period $\tau$, one finds the rectified current
\begin{equation}
  \Irect = {\omega \over 2 \pi}
  R \int_S dX_1 dX_2  \left(
  C_2 {\partial G \over \partial X_1} - C_1 {\partial G \over
  \partial X_2} \right).
  \label{eq:V}
\end{equation}
This expression has to be compared with the expression for the pumped
current in the case of adiabatic pumping,\cite{brouwer} which has the
same integration domain as Eq.\ (\ref{eq:I}), but a different
integrand.

For a quantum dot with an irregular shape (a
chaotic quantum dot), as in
Ref.\ \onlinecite{switkes}, the statistical distribution and the
magnetic field dependence of the 
conductance $G$ and its derivatives $\partial G/\partial X_1$,
$\partial G/\partial X_2$ are known in the literature.
The reader is referred to Ref.\ \onlinecite{beenakker} for a 
review. Here
we limit ourselves to a summary of the key qualitative properties 
of $\Vrect$ that can be compared to the experimental observations 
reported in Ref.\ \onlinecite{switkes}.
(The current $\Irect$ has the same qualitative properties.)

{\em 1.}
For small driving amplitudes $\delta X_1 = \delta X_2 \equiv \delta X$, 
one has $\Vrect
\propto (\delta X)^2$; For large $\delta X$, one has $\Vrect 
\propto (\delta X)^{1/2}$.
The crossover between small-amplitude driving and large-amplitude
driving occurs at amplitude $\delta X \sim (e^2/h)
/\sigma(\partial G/\partial X)$, corresponding to 
$\Vrect \sim (e^2/h)^2 G^{-2}
C \omega/\sigma(\partial G/\partial X)$, where
$\sigma(\ldots)$ denotes standard deviation with respect to
mesoscopic fluctuations.\cite{foot}

{\em 2.} 
The ensemble average $\langle \Vrect
\rangle = 0$.

{\em 3.} The dc\ voltage $\Vrect$ is symmetric
under reversal of the magnetic field, $\Vrect(B) = \Vrect(-B)$. 

{\em 4.} For multichannel point contacts, the variance of $\Vrect$
is decreased by a factor two when time-reversal symmetry is broken by
a magnetic field. (The experiment is for two-channel point 
contacts.) 

All of the properties 1 -- 4 were observed by Switkes {\em et al.}
Properties 1 and 2 also apply to the voltage
generated by an adiabatic quantum pump,\cite{brouwer,zhou,shutenko}
except for the magnitude of the voltage
at the crossover between small-amplitude and large-amplitude driving,
which is $\Vpump \sim \omega e/G$ for an adiabatic quantum pump,
corresponding to a current of $\sim 1$ electron per cycle.\cite{brouwer}
The dc\ voltage observed in Ref.\
\onlinecite{switkes} corresponds to $\sim 20$ electrons per
cycle at the crossover between small-amplitude and large-amplitude 
driving.
Properties 3 and 4 do not hold for an adiabatic quantum pump. 
In view of the magnetic field symmetry and the magnitude
of the observed dc\
voltage, it cannot be excluded that the voltage $\bar V$ measured in
Ref.\ \onlinecite{switkes}
is dominated by the rectification voltage $\Vrect$. 

How can the rectification voltage (\ref{eq:I})
be distinguished from the voltage generated by a true quantum pump
if, in principle,
both are present in the same experiment? 
First, when the rectification voltage is dominant, Eq.\ (\ref{eq:I})
should provide a verifiable relation between the measured dc\
voltage $\bar V$ and the conductance $G$ and its derivatives 
$\partial G/\partial X_{1,2}$, which can be measured 
independently of $\bar I$. Second, the two voltages are easily
distinguished by their magnetic field symmetry: the rectification
voltage is magnetic-field symmetric, while the voltage generated
by a quantum pump is not.  
It may be easier to measure pumping in a current
measurement geometry with low-resistance contacts, as, 
in that case, the rectification current scales
with the resistance $R$ of the circuit path containing the
current meter.

I would like to thank C.\ M.\ Marcus, D.\ C.\ Ralph, and M.\ G.\
Vavilov
for very helpful discussions. This work was supported by the
NSF under grant no.\ DMR 0086509 and by the Sloan foundation.


\begin{references}
\bibitem{foot} In Ref.\ \onlinecite{switkes}, the measured voltage
at the crossover between small-amplitude driving and large-amplitude
driving is 0.5 $\mu$V at driving frequency 10 MHz. 
The amplitude of the gate voltages at that point is 80 mV.
For comparison, the dc\ voltage corresponding to one electron 
per cycle at a driving frequency of 10 MHz is 0.02 $\mu$V.
\end{references}
\end{document}